\documentclass[traditabstract]{aa}

\usepackage{epsfig}
\usepackage{graphicx}
\usepackage[latin1]{inputenc}
\usepackage{txfonts}
\usepackage{rotating}
\usepackage{natbib}

\begin{document}

   \title{Spectral monitoring of RX J1856.5-3754 with XMM-Newton}

   \subtitle{Analysis of EPIC-pn data}

   \author{N. Sartore\inst{1},
A. Tiengo\inst{1,2},
S. Mereghetti\inst{1},
A. De Luca\inst{1,2,3},
R. Turolla\inst{4,5},
and F. Haberl\inst{6}
          }

   \institute{INAF - Istituto di Fisica Spaziale e Fisica Cosmica, via E. Bassini 15, 20133, Milano, Italy
              \email{sartore@iasf-milano.inaf.it}
	 \and
IUSS - Istituto Universitario di Studi Superiori, viale Lungo Ticino Sforza 56, 27100, Pavia, Italy
	 \and
INFN - Istituto Nazionale di Fisica Nucleare, sezione di Pavia, via A. Bassi 6, 27100, Pavia, Italy
	 \and
Dipartimento di Fisica e Astronomia, Universit\`{a} di Padova, via Marzolo 8, 35131, Padova, Italy
	 \and
Mullard Space Science Laboratory, University College London, Holmbury St. Mary, Dorking, Surrey, RH5 6NT, UK
	 \and
Max-Planck-Institut f{\"u}r extraterrestrische Physik, Giessenbachstra{\ss}e, 85748 Garching bei M{\"u}nchen, Germany
             }

   \date{Received ...; accepted ...}

\abstract{Using a large set of XMM-Newton observations we searched for long term spectral and flux variability of the isolated 
neutron star RX J1856.5-3754 in the time interval from April 2002 to October 2011. This is the brightest and most extensively observed 
source of a small group of nearby, thermally emitting isolated neutron stars, of which at least one member 
(RX J0720.4-3125, see e.g. Hohle et al., 2010) has shown long term variability. A detailed analysis of the data obtained with 
the EPIC-pn camera in the $0.15-1.2\,\rm keV$ energy range reveals only small variations  in the temperature derived with a single 
blackbody fit (of the order of $\sim 1\%$ around an average value of $kT^\infty\sim61\,\rm eV$). Such variations appear 
to be correlated with the position of the source on the detector and can be ascribed to an instrumental effect, most likely a spatial 
dependence of the channel to energy relation. For the sampled instrumental coordinates, we quantify this effect as variations 
of $\sim4\%$  and $\sim15\,\rm eV$ in the gain slope and offset, respectively. Selecting only a homogeneous subset of observations, 
with the source imaged at the same detector position, we find no evidence for spectral or flux variations of RX J1856.5-3754 
from March 2005 to present-day, with limits of $\Delta kT^\infty<0.5\%$  and  $\Delta f_X<3\%$ ($0.15-1.2\,\rm keV$ range), 
with $3\sigma$ confidence. A slightly higher temperature ($kT^\infty\sim61.5\,\rm eV$, compared to the average value 
$kT^\infty\sim61\,\rm eV$) was instead measured in April 2002. If this difference is not of instrumental origin, it implies a rate of 
variation $\sim -0.15\,\rm eV\,yr^{-1}$ between April 2002 and March 2005. The high-statistics source spectrum from the selected 
observations is best fitted with the sum of two  blackbody models, with temperatures 
$kT_{h}^\infty = 62.4_{-0.4}^{+0.6}\, \rm eV$ and $kT_{s}^\infty = 38.9_{-2.9}^{+4.9}\, \rm eV$, which can also account for the flux 
seen in the optical band. No significant  narrow or broad spectral features are detected, with upper limits of $\sim 6\,\rm eV$ 
on their equivalent width.}

   \keywords{Stars: neutron, Stars: Individual: RX J1856.5-3754}

\authorrunning{Sartore et al.}

\maketitle

\section{INTRODUCTION}

RX J1856.5-3754 (J1856 hereafter) is the prototype of the so-called XDINSs, i.e. the seven X-ray Dim Isolated Neutron Stars discovered 
by the ROSAT satellite (e.g. \citealt[see also]{wa1996} \citealt{ha2007, tu2009} for reviews). 
These are nearby, $d\lesssim300\,\rm pc$, radio-quiet isolated neutron stars characterized by thermal spectra, 
with temperatures\footnote{measured by an observer at infinity.} $kT^\infty\sim50-100\,\rm eV$ and luminosities 
$L_X\sim10^{31}-10^{32}\,\rm erg\,s^{-1}$. Six out of the seven XDINSs exhibit broad spectral features at energies between 270 and 700 eV, 
with equivalent widths of several tens of eV, interpreted as proton cyclotron lines or atomic transitions in strong magnetic fields, 
$B\sim10^{13}\,\rm G$. Spin periods of the XDINSs are in the 3--12 s range and, assuming magneto-dipole braking, 
their measured spin-down rates, of $\sim10^{-14}-10^{-13}\,\rm s\,s^{-1}$, imply magnetic fields $B\sim10^{13}-10^{14}\, \rm G$, 
in broad agreement with those inferred from the spectral features. All XDINSs have very faint optical/UV counterparts, $m_V\sim26-27$: 
the optical/UV flux exceeds that expected from the Rayleigh-Jeans tail of the X-ray blackbody \citep[BB, see e.g.][]{ka2011}.

Intriguingly, the second brightest XDINS, RX J0720.4-3125 (J0720 hereafter) showed significant long term variations of its spectral 
properties \citep[][and references therein]{ho2010}. The nature of these changes is still under debate and  can possibly be related 
to a precession of its spin axis \citep[e.g.][]{ha2006} or to a glitch-like episode \citep{vkk2007}. 
Another interesting fact is that XDINSs share some of their properties with magnetars. In particular, their spin periods fall in the same 
range, while the magnetic fields of XDINSs are intermediate between those of magnetars, $B\sim10^{14}-10^{15}\, \rm G$, 
and normal pulsars, $B\sim10^{11}-10^{13}\,\rm G$. Also, XDINSs seem to be older than magnetars, with spindown ages of 
$\sim10^5-10^6$ years against $\sim10^3-10^5$ years \citep[e.g.][]{me2008}. 
These similarities suggest a possible evolutionary link between the two classes of isolated neutron stars (INSs).

J1856 has the lowest pulsed fraction, $\sim 1\%$ \citep{tm2007}, and the weakest magnetic field, $B\sim1.5\times10^{13}\,\rm G$ 
\citep{vkk2008} among known XDINSs. Being the brightest, $f_X\sim1.5\times\,10^{-11}\,\rm erg\,s^{-1}\,cm^{-2}$ and closest 
of the category, $d=123_{-15}^{+11}\,\rm pc$ \citep[e.g.][]{walter2010}, it stands among the best candidates to probe the mass 
and radius of a neutron star, which in turn would be of paramount importance to constrain the equation of state (EOS) of 
matter at nuclear densities. Analyzing the data from a 57 ks XMM-Newton observation and a $\sim500$ ks observation by {\it Chandra}, 
\cite{bu2003} suggested that atmosphere models with heavy elements can be ruled out, given the lack of any spectral features. 
Also, they ruled out non-magnetic and fully ionized hydrogen atmosphere models, since both  would over-predict the optical flux. 
On the other hand, they found that the broadband (optical + X-ray) spectrum is well fitted by a two component BB, where the hotter 
component, $kT_X^\infty\simeq\,63.5\,\rm eV$ $(R_{X}^\infty\simeq\,4.4\,(d/120\,\rm pc)\,\rm km)$, is responsible for the X-ray 
emission while the cold component, $kT_{opt}\,<\,33\,\rm eV$ $(R_{opt}^\infty>17\,(d/120\,\rm pc)\,\rm km)$, accounts only for the 
optical emission. Alternative models, like emission from condensed matter surface \citep{la2001, tu2004}, or partially ionized 
hydrogen atmospheres \citep{ho2007} cannot be ruled out.

Thanks to its bright and presumably steady emission, J1856 has been targeted routinely for calibration purposes by XMM-Newton 
in the last decade. The large amount of collected data allows us to precisely characterize its spectral evolution on time scales 
from months to $\sim10\,\rm years$. At the same time, the stability of the detectors on board XMM-Newton in the same time-frame 
can be scrutinized as well, as done with the data taken before 2006 in \cite{ha2007}. 
Here we report on the analysis of all XMM-Newton observations of J1856 performed so far in imaging mode with the EPIC-pn camera 
\citep{st2001} on board XMM-Newton. The reduction and the analysis of the data is described in Section \ref{sect-data}. 
Results are also illustrated in Section \ref{sect-data}. We constrain the detector stability and obtain an upper limit for the 
spectral variations of the star and compare it with those of other isolated neutron stars reported in the literature. 
A discussion on the astrophysical implications of these results is also given (Section \ref{sect-disc}).

\section{DATA REDUCTION AND SPECTRAL ANALYSIS}\label{sect-data}

\begin{table*}[ht]
\centering
\caption{Log of all EPIC-pn observations of RX J1856.5-3754 performed in Small Window Mode.}
\begin{tabular}{ c c c c c }
\hline \hline
\\
Observation & Date & Obs.ID & Net Exposure Time & Counts \\
 & & & $[\rm s]$ & \\
\hline
A & 2002-04-08 & 0106260101 & 40030 & 302601 \\
B & 2004-09-24 & 0165971601 & 22960 & 174955 \\
C & 2005-03-23 & 0165971901 & 14190 & 106949 \\
D & 2005-09-24 & 0165972001 & 22700 & 168609 \\
E & 2006-03-26 & 0165972101 & 48340 & 361845 \\
F & 2006-10-24 & 0412600101 & 49820 & 374018 \\
G & 2007-03-14 & 0412600201 & 26940 & 204513 \\
H & 2007-03-25 & 0415180101 & 16540 & 126883 \\
I & 2007-10-04 & 0412600301 & 23920 & 179449 \\
J & 2008-03-13 & 0412600401 & 33080 & 245923 \\
K & 2008-10-04 & 0412600601 & 43580 & 330553 \\
L & 2009-03-19 & 0412600701 & 47510 & 353698 \\
M & 2009-10-07 & 0412600801 & 40780 & 306449 \\
N & 2010-03-22 & 0412600901 & 48380 & 360306 \\
O & 2010-09-29 & 0412601101 & 47870 & 356580 \\
P & 2011-03-14 & 0412601301 & 45210 & 342204 \\
Q1 & 2011-10-05* & 0412601501 & 17490 & 134747 \\
Q2 & '' & '' & 16310 & 124431 \\
Q2 & '' & '' & 16800 & 127176 \\
Q4 & '' & '' & 17920 & 128549 \\
\hline
\multicolumn{4}{l}{(*) The observation of October 2011 has been divided in 4 pointings,}\\
\multicolumn{4}{l}{with the source at different positions on the detector.}\\
\end{tabular}
\label{obs-log}
\end{table*}

We use all imaging data collected with the EPIC-pn camera from April 2002 to October 2011. Data from the two EPIC-MOS detectors 
are not considered in this analysis because the MOS effective area at soft X-ray energies is much smaller than that of the pn and 
the MOS cameras are known to be less stable on long term \citep{re2006}. Table \ref{obs-log} reports the date, identification number, 
net exposure time, and number of background-subtracted counts of J1856 for each observation. All observations were performed in Small Window mode 
with the thin filter. Raw data are processed with the XMM-Newton science analysis package (SAS) v11.0, using the {\tt epproc} pipeline. 
We then build light curves above 10 keV, with bin size of 100 s, in order to identify soft proton flares. Based on these light curves, 
we clean the data by removing time intervals with count rate higher than $4\sigma$ above the mean rate. Source spectra 
are then extracted from a circular region of 30" radius, selecting only single pixel events (i.e. PATTERN=0 in the {\tt xmmselect} task), 
in order to obtain event lists of the highest possible quality. The extracted spectra are binned in order to have at least 30 counts 
per bin. The spectral analysis is performed with XSPEC 12 \citep{ar1996}, selecting photon energies in the $0.15-1.2$ keV range. 

\begin{figure*}[b]
\includegraphics[width=0.8\textwidth]{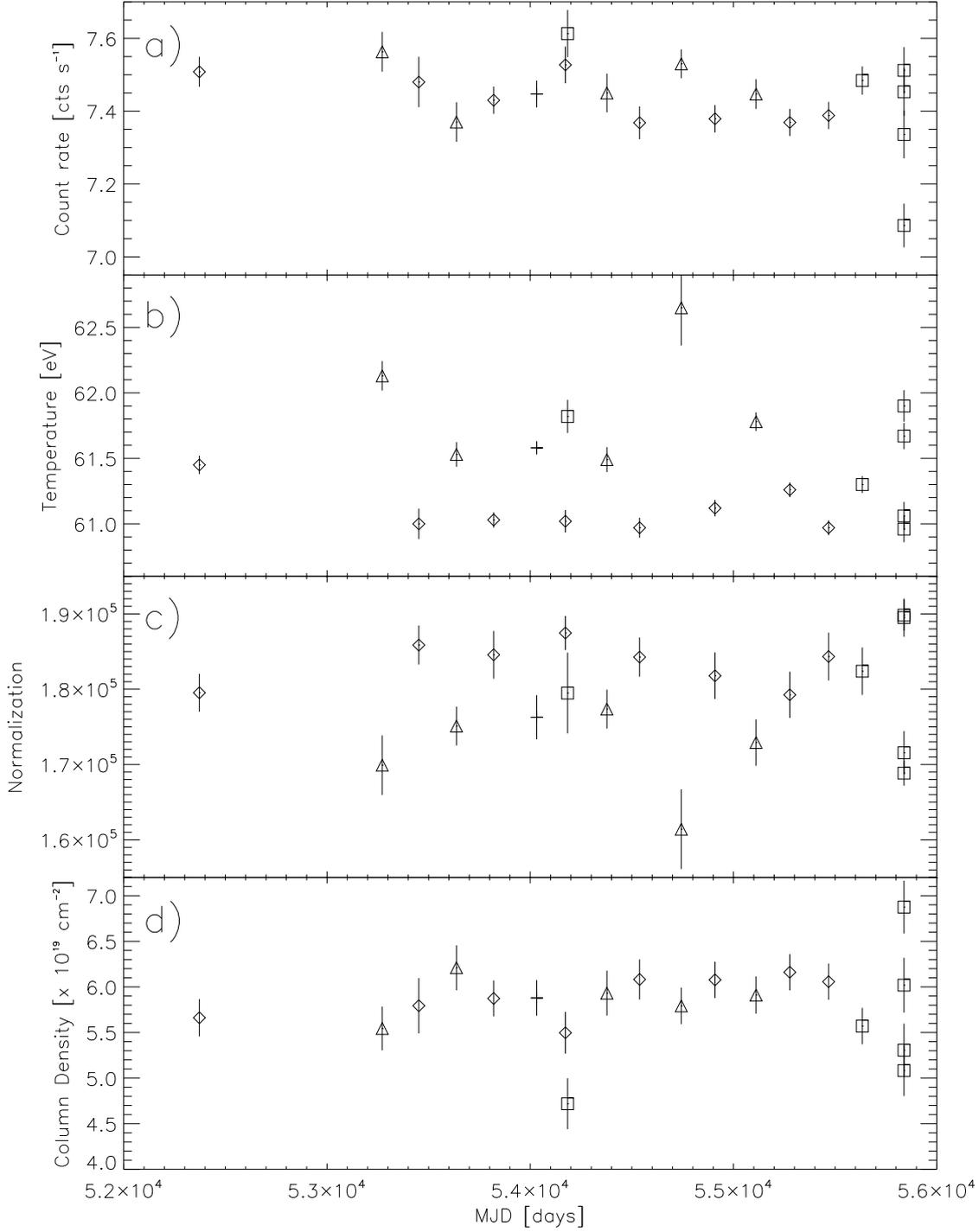}
\caption{Long term evolution of the spectral parameters obtained with a single BB fit in the  $0.15-1.2$ keV range.
Panels \textit{a}, \textit{b}, and \textit{c}, show the count rate, temperature, and BB  normalization of a simultaneous fit with the 
column density fixed to a common value. Panel \textit{d} indicates the  column density obtained when, instead, the BB parameters are 
fixed to a common value for all the observations (see text).  Diamonds, triangles and squares 
represent `soft', `hard' and off-axis observations, respectively. The plus symbol is the October 2006 observation, which is at an 
intermediate detector position (see Fig. \ref{fig-detector}). Error bars correspond to $3\sigma$ 
confidence intervals.}
\label{fig-evo}
\end{figure*}

\begin{figure*}[b]
\includegraphics[width=0.9\textwidth]{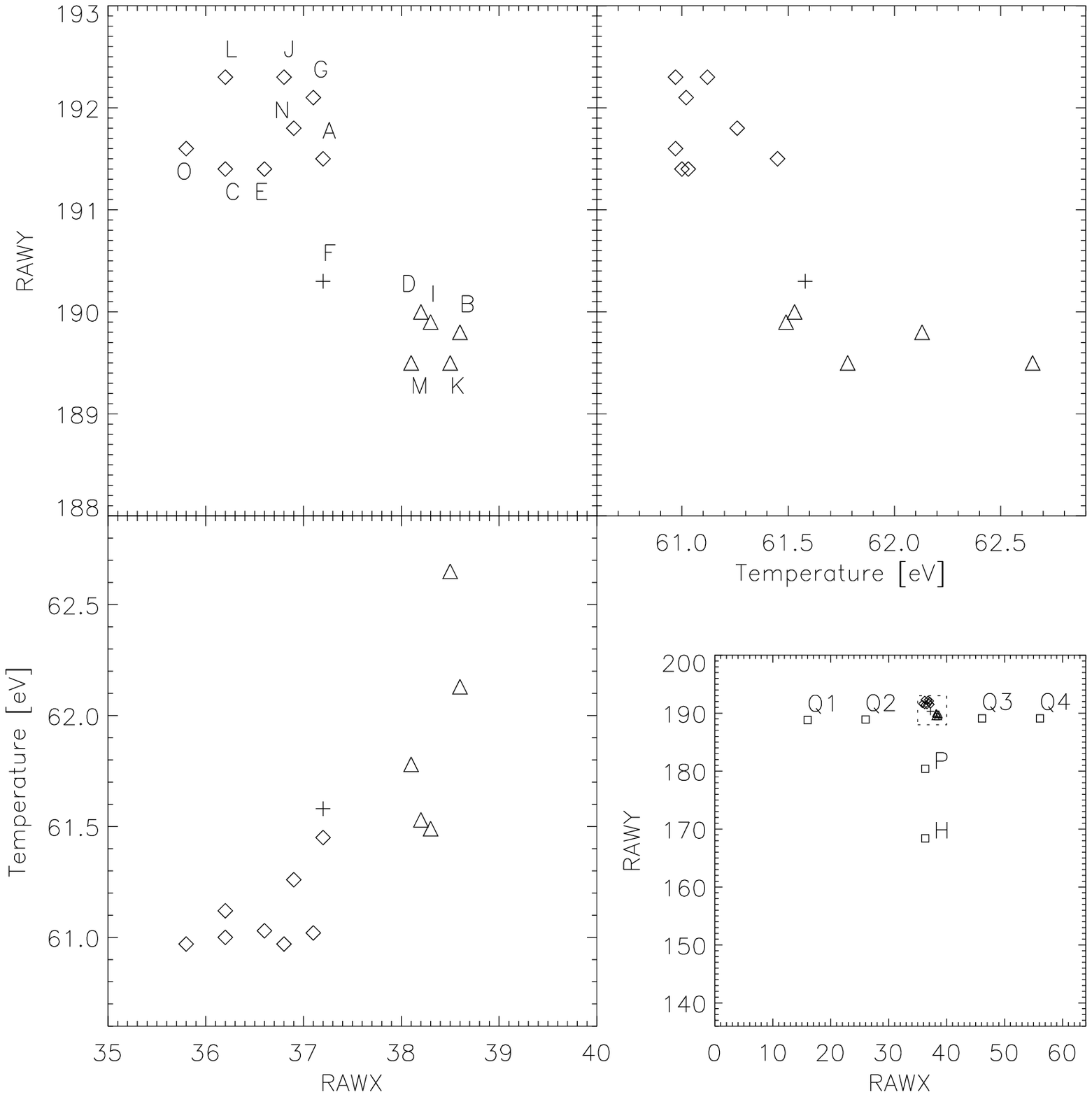}
\caption{Lower right panel: position of the source in detector coordinates (RAWX,RAWY). The region in the dashed square is enlarged 
in the upper left panel. The symbols indicate the different observations: `soft' region (diamonds), `hard' region (triangles), 
observation F (plus), off-axis observations (squares). The other two panels indicate the temperature as a function of the 
source position.}
\label{fig-detector}
\end{figure*}

We perform a simultaneous fit of all the 20 spectra with a single absorbed BB, using a \texttt{phabs*bbodyrad} model 
and adopting the abundances of \cite{ag1989}. The true spectra of XDINSs can be more complex than a single BB. 
However, we are interested in relative variations in the emission of J1856. Thus, we rely on the parameters of 
the single BB in order to quantify these variations. The best fit gives a BB temperature $kT^\infty=61.30\pm0.04\,\rm eV$, 
and a column density $N_{\rm H}=(5.84\pm0.04)\times10^{19}\,\rm cm^{-2}$ ($\chi_\nu^2=1.78$ with 2784 degrees of freedom). 
The residuals show large systematic deviations, especially above $\sim 0.5\,\rm keV$ (Fig. \ref{fig-spec-1bb}, upper panel), 
which change from observation to observation, indicating that some of the spectral parameters vary between different observations. 
Therefore, we repeat the spectral fit leaving the BB temperature and normalization as free parameters for each observation. 
In this case we obtain a better $\chi_\nu^2=1.37$ with 2746 degrees of freedom. The resulting parameters are shown in Fig. \ref{fig-evo} 
(panels \textit{a}, \textit{b}, \textit{c}). The fit of the 20 spectra with the column density as free parameter, 
and a unique temperature and normalization for all observations, is not as good as the previous one, $\chi_\nu^2=1.58$ with 
2765 degrees of freedom. The resulting column densities are shown in Fig. \ref{fig-evo} (panel \textit{d}). 
The count rate shows variations of $\sim1-2$ percent around a mean of $\sim 7.46\,\rm cts\,s^{-1}$, but with no evidence 
of a defined trend and without correlation with the BB temperature or normalization. On the other hand, the hydrogen column density 
is clearly anti-correlated with the count rate (compare panels \textit{a} and \textit{d} in Fig. \ref{fig-evo}), 
indicating that the changes in the count rate are mainly determined by differences in the softest spectral band. 
As already noticed by \cite{stu2010}, the temperature also shows variations of $\sim 1-2$ percent ($\sim1.5-2\,\rm eV$), 
which are anti-correlated with the normalization. We note that these variations have a seasonal pattern with a period of 
$\sim12$ months, suggesting that they are likely related to instrumental effects rather than being intrinsic changes of the source 
emission. To verify this hypothesis we study the relation of the BB temperature with the position of the source centroid 
on the detector (RAWX and RAWY coordinates). With the exception of the three off-axis observations (observations H, P and Q, 
which includes the four separate pointings performed in October 2011) and one in an intermediate position (observation F), 
in most of the observations the source centroid lies in two close but distinct regions of the detector (Fig. \ref{fig-detector}), 
one at RAWX$\sim$36-37, RAWY$\sim$192, and the other one at RAWX$\sim$38, RAWY$\sim$190 (hereafter `soft' and `hard' region, respectively, 
according to the relative hardness of their spectra, as outlined below). This is due to the different roll angle of the satellite in the 
Spring and Fall visibility intervals for the J1856 sky region. 

The best fit BB temperature is systematically higher for observations with the source in the `hard' region 
(upper right and lower left panels of Fig. \ref{fig-detector}), and in the 3 most extreme off axis observations at RAWY$=$168, 
RAWX$=$16 and RAWX$=$56. In the pn camera the signal is read independently from each column along the RAWY position. 
This implies that small inaccuracies in either the gain or the charge transfer inefficiency (CTI) correction along the readout direction 
could be the underlying cause for the observed spectral differences. Thus we fit again all the spectra with a unique BB model, 
but this time allowing a variation of the gain parameters. These parameters, \textit{slope} and \textit{offset}, 
define the relation between the energy scale in the response matrix of each observation with respect to that of an observation taken 
as reference: $E_i'=E/slope_i-offset_i$. We assume the spectrum of the first observation (observation A) as reference, 
i.e. we fix its \textit{slope} and \textit{offset} at 1 and 0, respectively. In this case we obtain $\chi_\nu^2=1.32$ 
with 2746 degrees of freedom, with a net reduction of the scatter of fit residuals for all spectra (Fig. \ref{fig-spec-1bb}, lower panel) 
and values of the gain parameters ranging from $\sim0.986$ to $\sim1.028$ for the \textit{slope} and from $\sim-6.6$ eV to $\sim8.1$ eV 
for the \textit{offset}. This implies that there could be variations up to $\sim4\%$ of the \textit{slope} and of 
$\sim15\,\rm eV$ of the \textit{offset}\footnote{Note that the PN channel energy has a width of $\sim5$ eV, which means that the photon 
energy as detected by the instrument is not known better than that.} 
throughout the $\sim10$ years and source positions sampled by the XMM-Newton observations 
(see Fig. \ref{fig-detector}). This is a conservative estimate since we considered the two most extreme spectra and assumed no intrinsic 
spectral variability in J1856.

\begin{figure*}[b]
\includegraphics[width=0.7\textwidth,angle=270]{spec-single-bb.ps}
\includegraphics[width=0.7\textwidth,angle=270]{spec-single-bb-gain.ps}
\caption{Simultaneous fit of all J1856 spectra with the same single BB. Upper (lower) panel shows the residuals without (with) 
the gain parameters free to vary.}
\label{fig-spec-1bb}
\end{figure*}

\subsection{CONSTRAINTS ON INTRINSIC SPECTRAL VARIATIONS}\label{cooling-trend}

To investigate possible variations of the spectral properties of J1856 during the nine year-long time-span of the monitoring campaign, 
we have to compare only homogeneous data sets, i.e. when the source was located at approximately the same detector coordinates, 
in order to reduce the incidence of the systematic uncertainty reported in the previous section.

\begin{figure*}[b]
\includegraphics[width=0.7\textwidth]{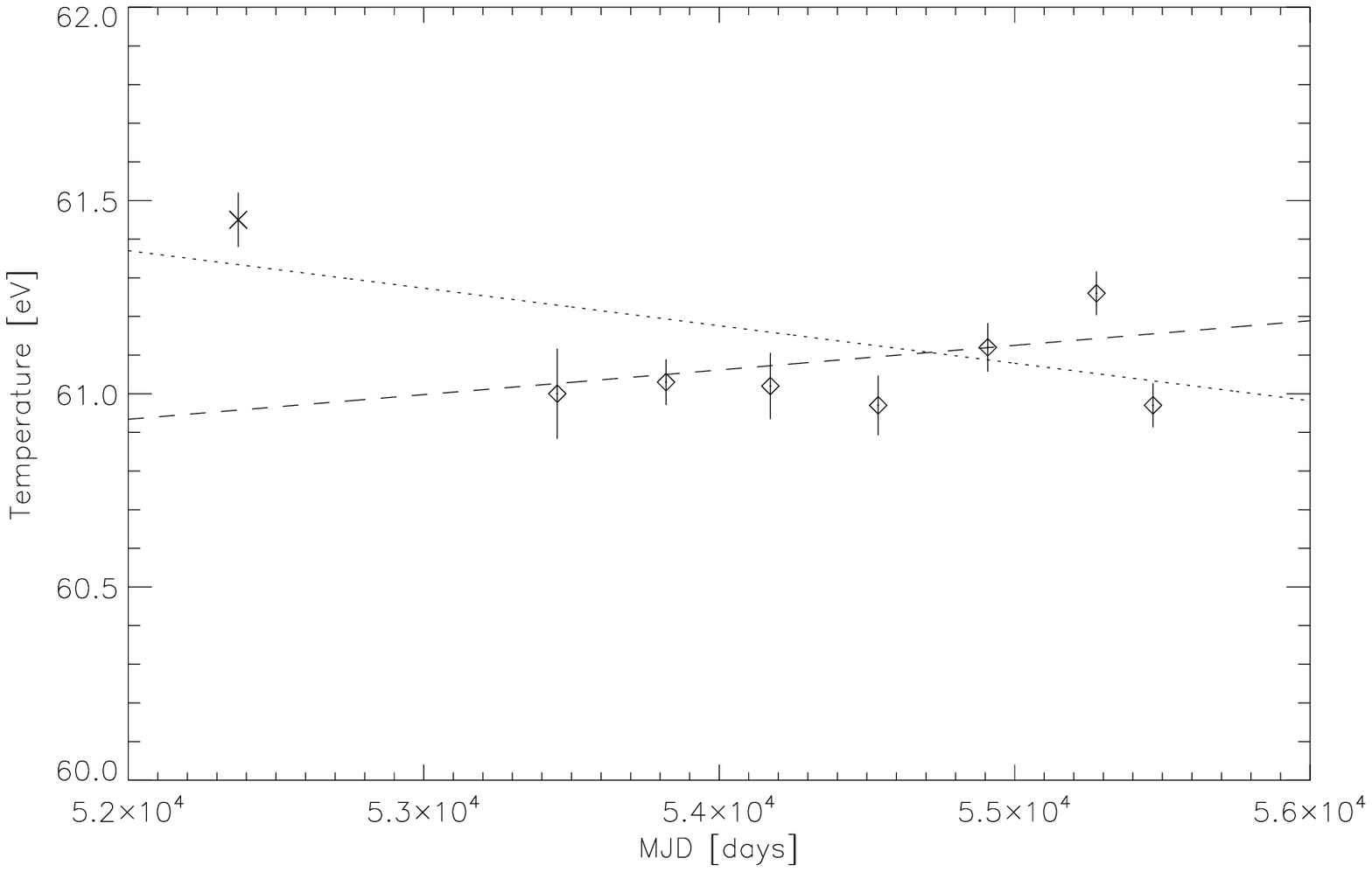}
\caption{Long term evolution of the temperature for `soft' observations. The dotted line represents the linear fit to the data. 
The dashed line represents a linear fit to the data without considering the observation of April 2002, which is marked by the X symbol. 
Error bars correspond to $3\sigma$ confidence intervals.}
\label{fig-kt-trend}
\end{figure*}

First, we study temperature variations considering the 8 spectra of the observations with the source in the `soft' region. 
Fitting a linear function to the derived temperatures, we obtain a negative (cooling) rate of variation of 
$\sim-0.035\pm0.010\,\rm eV\,yr^{-1}$ (Fig. \ref{fig-kt-trend}). However, the fit is unacceptable 
($\chi_\nu^2=5.16$ for 6 degrees of freedom) because of the first data point (observation A), which differs significantly 
from the others. Assuming no variations of the instrument response between April 2002 and March 2005, the `anomalous' temperature 
inferred from the older spectrum may be explained by an intrinsic change of the source emission, like those observed in J0720, 
implying a relatively rapid drop in the temperature, of $\sim0.5$ eV in less than 3 years. In alternative, some subtle variations 
of the instrument (energy response of the pixel column) may have occurred in the same time-frame. Excluding the first data point 
we obtain a better fit ($\chi_\nu^2=1.74$ for 5 degrees of freedom) and the rate of temperature variations of 
$\sim0.023\pm 0.015\,\rm eV\,yr^{-1}$, which is \textit{de facto} consistent with a constant temperature at $2\sigma$ level 
(Fig. \ref{fig-kt-trend}).

If instead we use the data from the 5 `hard' observations, we obtain a rate of temperature variation of 
$\sim0.044\pm0.026\, \rm eV\,yr^{-1}$, again consistent with a constant temperature at $2\sigma$ level. However the value obtained 
is affected by larger uncertainty since, as it can be seen from the middle panel of Fig. \ref{fig-evo}, in this case there is 
a larger scatter of the $kT^\infty$ values with respect to those found for the `soft' observations. The origin of this 
larger scatter is unclear and could be possibly related to a gain instability of the pixel columns within the source 
point spread function. In any case, owing to their homogeneity and longer total exposure, throughout the rest of the paper 
we will make use of only the `soft' data in order to constrain the spectral properties of J1856.

\subsection{BEYOND THE SINGLE BLACKBODY}\label{sect-spec}

In order to obtain a single spectrum with high statistics, we merge together the 7 spectra of the `soft' group (Fig. \ref{fig-detector}) 
excluding, as before, the anomalous observation of April 2002 (observation A). The resulting spectrum contains 
$\sim1.9\times10^6$ background-subtracted counts and corresponds to a net integration time of $\sim254$ kiloseconds.

\begin{figure*}[b]
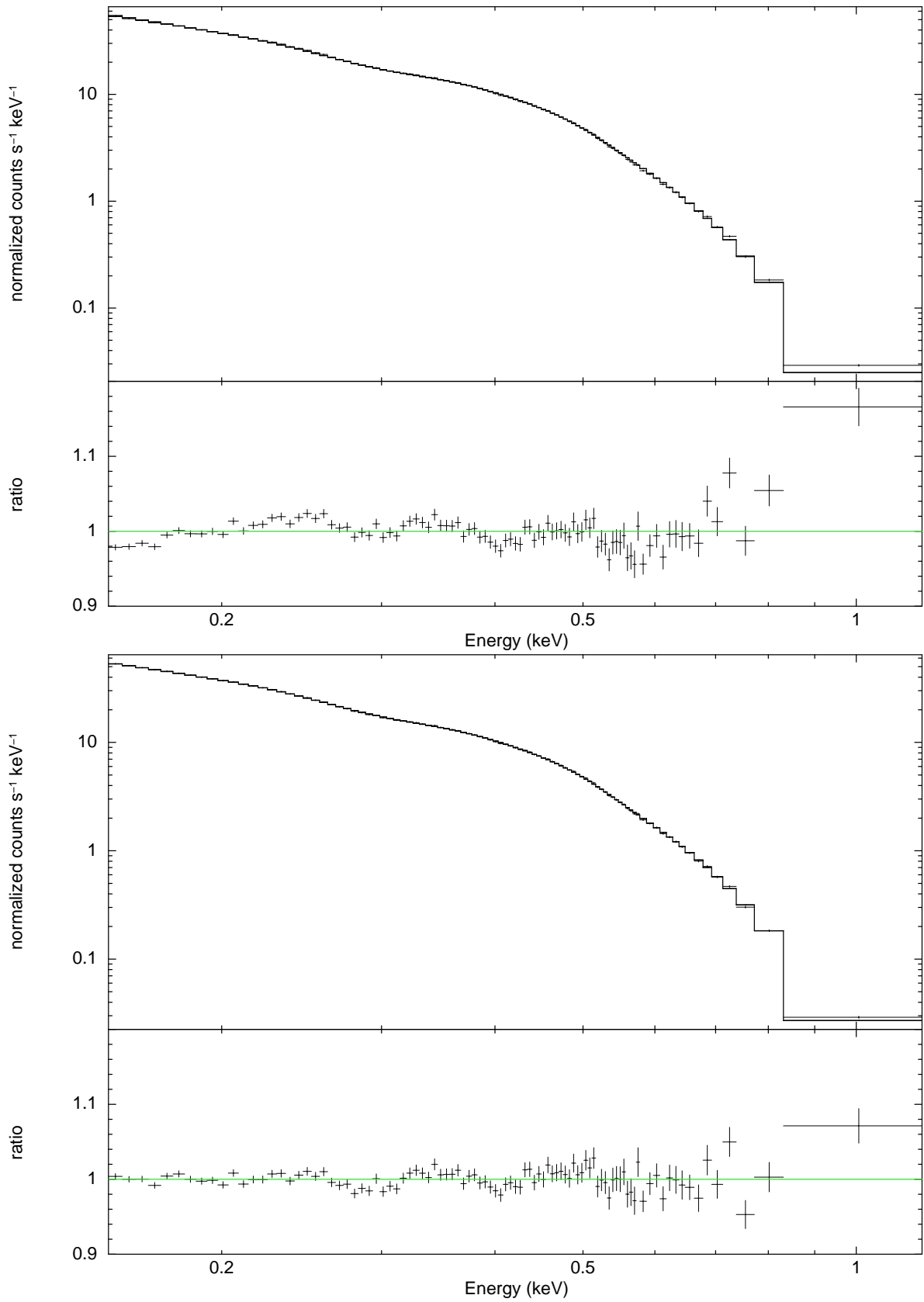

\includegraphics[width=0.6\textwidth,angle=270]{spec-1bb-sys15.ps}
\includegraphics[width=0.6\textwidth,angle=270]{spec-2bb-sys06.ps}
\caption{(Top) Single BB fit of the merged spectrum obtained from observations of the 'soft' group. 
(Bottom) Same as Top panel but with a two BB model.}
\label{fig-joint-spec}
\end{figure*}

We first try fits with a single BB (Fig. \ref{fig-joint-spec}, upper panel). In order to obtain a formally acceptable fit 
(null hypothesis probability $>0.1$) we add a systematic error of $1.5\%$ (\textit{systematic} parameter in XSPEC). 
This is an energy-independent systematic error added to the model in XSPEC and accounts for the uncertainties on the spectral model 
and for likely residual calibration inaccuracies. We obtain a reduced $\chi_\nu^2=1.12$ for 176 degrees of freedom. The resulting 
temperature is $kT^\infty=61.5\pm0.1\, \rm eV$, to which corresponds an emission radius $R^\infty=5.0\pm0.1\,\rm km$ for a distance 
of 120 pc. The column density is $N_{\rm H}=(4.8\pm0.2)\times10^{19}\, \rm cm^{-2}$. The fit residuals exhibit systematic deviations 
both at the lowest and highest ends of the energy range (Fig. \ref{fig-joint-spec}, upper panel), which may be indicative of 
the inability of the single temperature BB to adequately fit the spectrum of J1856.

\begin{figure*}[b]
\includegraphics[width=0.9\textwidth]{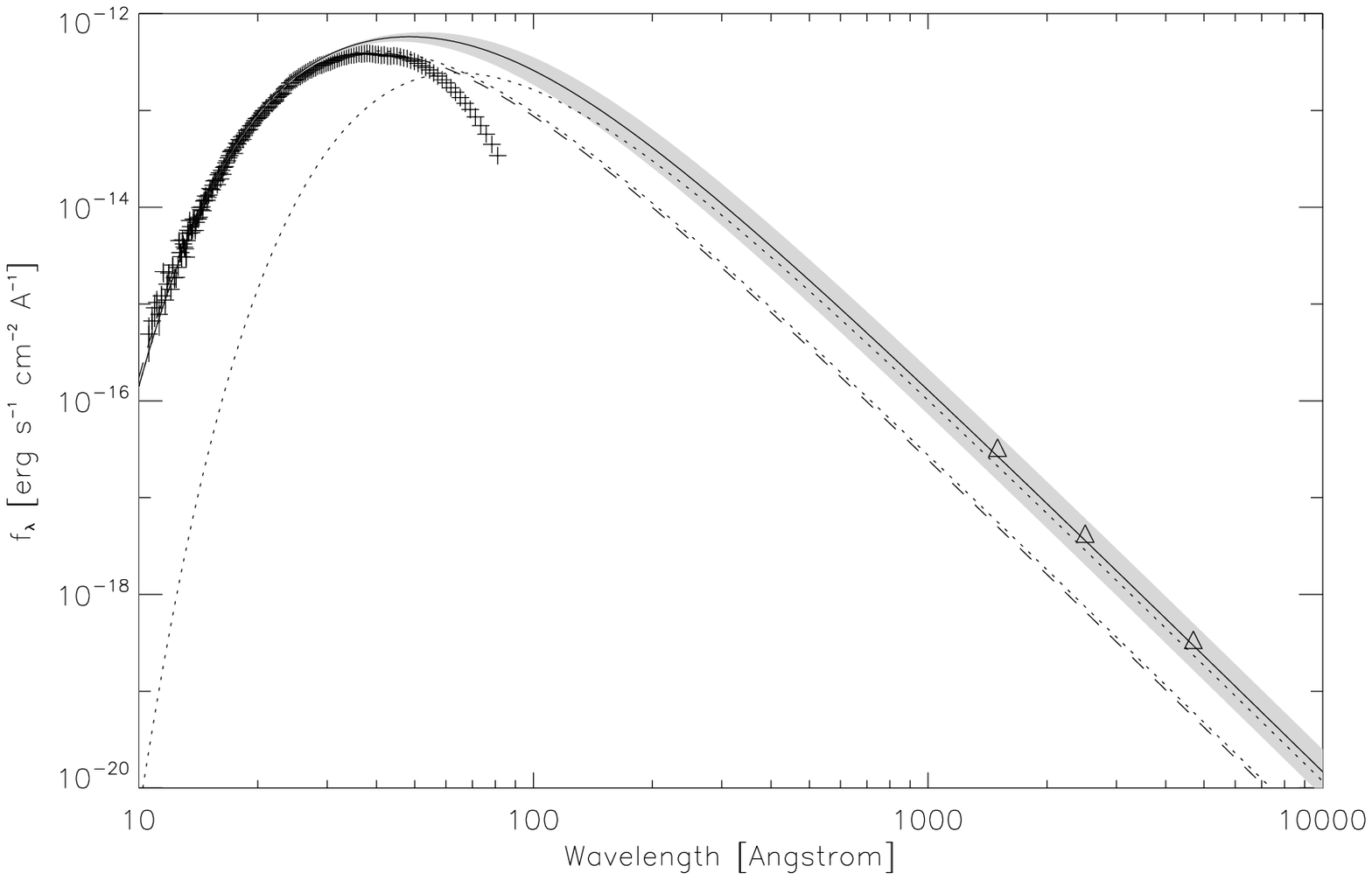}
\caption{Broadband spectrum of J1856. The dashed line is the extrapolation of the (unabsorbed) X-ray BB from \cite{bu2003}. 
The  dotted lines indicate the two  BB components obtained from the sum of all the homogeneous observations 
with the source in the `soft' region,  and the solid line is their sum. The shaded area marks the $1\sigma$ confidence region 
for the best fit model. The triangles represent the optical/UV data obtained from HST photometry \citep{ka2011}.}
\label{fig-broadband}
\end{figure*}

The presence of a second BB at different temperature is expected to explain pulsations \citep[e.g.][]{tm2007}. We therefore fit 
the merged spectrum with a two BB model (Fig. \ref{fig-joint-spec}, lower panel). In this case we obtain an acceptable fit by adding a 
systematic error of $0.6\%$, which returns a $\chi_\nu^2=1.11$. The column density is now 
$N_{\rm H}=(12.9\pm2.2)\times 10^{19}\,\rm cm^{-2}$. The hard BB has a temperature of $kT_h^\infty=62.4_{-0.4}^{+0.6}\,\rm eV$ 
with emission radius of $R_{h}^\infty=4.7_{-0.3}^{+0.2}\,(d/120\,\rm pc)\,km$, while the soft BB has a temperature 
$kT_s^\infty=38.9_{-2.9}^{+4.9}\, \rm eV$ and  emission radius $R_{s}^\infty=11.8_{-0.4}^{+5.0}\,(d/120\,\rm pc)\,km$. 
We summarize these results in Table \ref{tab-joint-spec}. Interestingly, the contribution of both the soft and hard X-ray blackbodies 
to the optical flux can account for the excess observed in J1856 \citep{vkk2001,ka2011}. Considering the best-fit parameters for 
the double BB model, the contribution of the soft BB at optical wavelengths is $\sim 4$ times larger than that of the hard BB. 
This implies an overall increase of a factor $\sim5$ with respect to the optical flux expected from the Rayleigh-Jeans tail of 
the hard BB alone (Fig. \ref{fig-broadband}). This value is consistent, within uncertainties, to the factor $\sim7$ obtained from 
the comparison between optical/UV photometry and the extrapolation of the X-ray data \citep{bu2003,ka2011}. 

We note three dips at $\sim$ 0.3, 0.4 and 0.6 keV in the residuals of both single and two-component BB fits. 
Since their width is smaller than the energy resolution of the pn detector in the range of interest, they are most likely due to 
non perfect instrumental calibrations, as already suggested by \cite{ha2007}. We search for other possible features in the spectrum of 
J1856,  adopting the single BB model with no systematic errors in order to facilitate the comparison with the spectral features found 
in other XDINS. We investigate the presence of narrow absorption features from 300 to 700 eV at 100 eV intervals. 
We use the \texttt{gaussian} model from the XSPEC library, imposing fixed line energy and 
width\footnote{The condition $\sigma=0$ implies that the line width is smaller than the energy resolution. The width $\sigma$ is thus 
automatically adjusted to match the minimum width imposed by the instrument resolution.}
($\sigma=0$ in XSPEC) and negative normalization.
We find a feature at $400\,\rm eV$, most likely one of the aforementioned dips and therefore of instrumental origin, 
with equivalent width of $2.0\pm0.5\, \rm eV$ at $4\sigma$ confidence. No other narrow features are found at the same confidence level, 
corresponding to a maximum equivalent width of $\sim6\,\rm eV$. Typically, XDINSs exhibit broad absorption lines. Hence, we look for 
broad features (fixed line width $\sigma=0.1\,\rm keV$) but with line energy free to vary. We find a feature at $370\pm15\, \rm eV$ 
and equivalent width $9\pm2\,\rm eV$ at $4\sigma$ confidence. However, the reality of this line is debatable and could arise from 
an attempt to fit a non purely Planckian spectral continuum. In fact, if we adopt the two-component BB as starting model, 
no broad features are found at $4\sigma$ confidence level.

\begin{table}[b]
\centering
\caption{Summary of fit parameters for the joint spectrum.}
\begin{tabular}{ c c c}
\hline \hline
Parameter & Single BB & Two BB \\
\\
$N_H$ & $4.8_{-0.2}^{+0.2}$ & $12.9_{-2.3}^{+2.2}$ \\
$[10^{19}\,\rm cm^{-2}]$ & & \\
\\
$kT_h^\infty$ & $61.5_{-0.1}^{+0.1}$ & $62.4_{-0.4}^{+0.6}$ \\
$[\rm eV]$ & & \\
\\
$R_{h}^\infty$ & $5.0_{-0.1}^{+0.1}$ & $4.7_{-0.3}^{+0.2}$ \\
$[\rm km]$ & & \\
\\
$kT_s^\infty$ & - & $38.9_{-2.9}^{+4.9}$ \\
$[\rm eV]$ & & \\
\\
$R_{s}^\infty$ & - & $11.8_{-0.4}^{+5.0}$ \\
$[\rm km]$ & & \\
\\
$\sigma_{sys}$ & $1.5\%$ & $0.6\%$ \\
\\
$\chi^2_\nu$ & 1.12 & 1.11 \\
\hline
\end{tabular}
\label{tab-joint-spec}
\end{table}\bigskip

\section{DISCUSSION}\label{sect-disc}

A spectral analysis of all the imaging observations of the isolated neutron star RX J1856.5-3754, performed from 2002 to 2011 with 
the EPIC-pn instrument on-board the XMM-Newton satellite, revealed small amplitude variations, of the order of $\sim 1-2$\%, 
quantified by changes in the fit parameters obtained with a single absorbed BB model. The long term trend of these variations suggested an 
instrumental origin, since there is a correlation of the fit parameters with the position of the source image on the detector. 
This is likely related to a non-uniform energy response between different channels of the readout electronics. 
The related uncertainty has been quantified with a gain fit in XSPEC, resulting in variations of the gain \textit{slope} and 
\textit{offset} of  $\sim4\%$ and $\sim15\,\rm eV$, respectively, over the pn detector positions covered by the J1856 observations 
(Fig. \ref{fig-detector}).

Using only homogeneous data, we found that the rate of temperature variations, in the framework of the single BB model, 
is compatible with a constant at $2\sigma$ confidence level over the last $\sim$5 years of the monitoring campaign. 
A higher temperature was recorded on April 2002. If not caused by subtle alterations of the instrument response, 
this difference would imply that also J1856 undergoes spectral changes, albeit the magnitude of these changes is much smaller than 
that observed in J0720. The observed temperature changed of $\sim0.5$ eV in three years (or less, considering the lack of coverage 
between April 2002 and March 2005) corresponds to a rate of $\sim-0.15\,\rm eV\,yr^{-1}$. In any case, the long term behavior of 
J1856 is markedly different from that of J0720. The latter underwent substantial and continuous changes in its observed spectral 
properties during the years. Apart the slightly higher temperature recorded in April 2002, J1856 exhibits instead a steady behavior, 
with no spectral variations in the last 5 years.

By merging all the homogeneous data-sets we obtained a high statistics spectrum with which we tested the validity of single and 
two BB models and checked for the presence of features in the spectrum of J1856. Apart some narrow features likely due to 
calibration issues and already reported in the literature \citep{ha2007}, we found no convincing evidence for broad or narrow absorption 
features. The absence of lines thus remains one of the distinctive properties of J1856 among XDINSs \citep[see also][]{ho2011}. 
The two BB model returns a better fit to the data and is more physically justified by the observed pulsations. The presence of a second, 
cooler BB component was invoked by \cite{po2002} to explain the optical/UV data but its contribution to the X-ray flux was considered 
negligible because of the apparent lack of pulsations reported for J1856 at that time \citep{br2002, bu2003, tru2004}. 
This implied an upper limit on the temperature of the cold BB, $kT_{opt}^\infty<33\,\rm eV$, which in turn yielded a lower limit on the 
stellar radius, $R^\infty=((R_{opt}^\infty)^2+(R_X^\infty)^2)^{1/2}>16\,(d/120\,\rm pc)\, km$. 
An even larger value was found by \cite{ham2011} for RBS 1223 (1RXS J130848.6+212708), another XDINS. 
It must be noted however that they assumed a condensed surface with a thin hydrogen atmosphere model to fit the spectrum of this 
neutron star. Thus, the measure of the radius resulting from the fit is larger than that obtained with a BB spectrum. 
From our analysis we found instead a smaller radius, 
$R^\infty=((R_s^\infty)^2+(R_h^\infty)^2)^{1/2}=12.7_{-0.2}^{+4.6}\,(d/120\,\rm pc)\, km$, 
consistent at $1\sigma$ level with the lower limit reported by \citeauthor{tru2004} Thus, our result allows a wider range of radii and 
is not much constraining for the EOS. In any case, we stress that the inferred parameters of a second BB in the X-ray spectrum of 
J1856 must be taken with caution. In addition to the statistical and systematic errors reported above, the normalization and thus 
the radius of soft BB component is affected by systematic uncertainties because of the not well constrained energy redistribution 
below $\sim 0.4\, \rm keV$. Large systematic deviations are expected also because J1856 is one of the main calibration targets used to 
determine the instrumental energy response and so the pn response matrix is currently tuned to reproduce 
a X-ray spectrum of J1856 not in contrast with its optical/UV flux, if a double BB models is assumed.
With the caveat of the aforementioned uncertainties, the extrapolation of the cold BB accounts also for a large fraction of 
the reported optical excess \citep[e.g.][]{ka2011}, avoiding the need for more complex models proposed to explain the observed 
optical/UV flux. 

\section{CONCLUSIONS}\label{sect-conc}
The results presented in this work show that the small amplitude variations in the spectral parameters of the isolated neutron star 
RX J1856.5-3754, obtained by fitting its spectrum with a single absorbed BB model, are due to a non-uniform energy response 
of the EPIC-pn camera. Once this instrumental effect is taken into account, the upper limits on the relative temperature and 
flux variations in the period March 2005-present are $\Delta kT^\infty<0.5\%$  and $\Delta f_X<3\%$, respectively. 
These can be taken as a measure of the source+instrument long term stability. A higher temperature was instead observed in April 2002. 
If due to an intrinsic change of the source spectral properties, it would imply that variations on time-scales of years, like those 
observed in the other XDINS RX J0720.4-3125, might be a common feature of this class of sources. 

We point out that, being the two brightest and most observed XDINSs, a continuous monitoring of these two sources will help to 
characterize the differences and analogies in their long term behavior and thus will give insights about the physical conditions on 
their surfaces, like e.g. the magnetic field and temperature distributions. In the case of J1856, the steady emission allowed us to 
sum all the data taken with the source in the same position on the detector, in order to obtain a spectrum with high count statistics. 
The resulting spectrum is best fitted by a two BB model, which is also more justified than a single BB by the observation of 
pulsations at X-ray energies. Also, the extrapolation at optical wavelengths of the emission from the two BBs accounts 
for the excess reported in the literature. However, due to uncertainties in the calibration of the pn at low energies, 
the radius of the star is not well constrained.

Finally, the results presented here suggest that J1856 is possibly the best choice as calibration source for instruments observing in the 
soft X-ray domain. Moreover, the EPIC-pn detector on board XMM-Newton is extremely stable over a long time-span (see above) and so, 
after taking into account a subtle spatial dependence of its spectral response, it can be considered as a reference instrument for the 
study of long term spectral variability of X-ray sources.


\begin{acknowledgements}
We thank the anonymous referee for the useful comments which improved the previous version of the manuscript.
We also thank M. Guainazzi and K. Dennerl for carefully reading the manuscript and for helpful discussion and suggestions. 
The XMM-Newton project is an ESA Science Mission with instruments and contributions directly funded by ESA Member States and 
the USA (NASA). We acknowledge the support of ASI/INAF through grant I/009/10/0.
\end{acknowledgements}




\end{document}